\documentclass[aps,prl,twocolumn,superscriptaddress,groupedaddress]{revtex4}
\usepackage{subfig}
\usepackage{graphicx}  
\usepackage{dcolumn}   
\usepackage{bm}        
\usepackage{amssymb}   
\usepackage{slashed}
\usepackage{graphicx}				
\usepackage{amsmath}
\usepackage{mathtools}
\usepackage{tikz,pgf}
\usepackage{comment}
\usepackage{asymptote}
\usetikzlibrary{arrows,backgrounds}
\usetikzlibrary{fit,scopes,calc,matrix,positioning,decorations.pathmorphing}
\usepackage[all]{xy}
\usepackage{yfonts}

\begin{document}
\title{On the T-dual renormalisation of entanglement entropy}
\author{Andrei T. Patrascu}
\address{ELI-NP, Horia Hulubei National Institute for R\&D in Physics and Nuclear Engineering, 30 Reactorului St, Bucharest-Magurele, 077125, Romania}
\begin{abstract}
Imposing T-duality in the renormalisation process of entanglement entropy leads to new relations between entanglement entropy counter-terms. T-duality is made explicit by means of the generalised metric of double field theory in the context of bulk-boundary duality. Double field theory in the bulk naturally provides the new relations between higher order quantum corrections to entanglement entropy as well as a systematic approach to understanding entanglement entropy renormalisation counter-terms. An analogue for Slavnov-Taylor identities for T-dual counter-terms of entanglement entropy is envisaged. 
\end{abstract}
\maketitle
\section{1. Introduction} 
The extended nature of strings allows for phenomena that are not conceivable from the perspective of field theory based on point particles. Despite the fact that many such exclusively string phenomena have been described, the connection to low energy phenomenology has always been a weak point of string theory. It appears however that string geometry is related to low energy observations in more subtle ways [1]. One of the profoundly string-theoretical effects is the T-duality. It represents a connection between radically different geometries and topologies. Given a compact background, when a string is moving on it, aside the states associated to KK compact momenta, we may also have winding numbers corresponding to the string turning around the non-contractible cycles. This leads to new winding states with the winding number representing how many times the cycle is wrapped by the string. Therefore, in general, string quantum states are labeled both by specific values of KK momenta and by the associated windings. T-duality encodes the fact that two theories with momenta and winding numbers interchanged may be physically indiscernible. An example for such a situation is the compactification of the bosonic string on a circle of radius $R$ which provides us with a theory that is equivalent to a string theory compactified on a circle of radius $\tilde{R}=\frac{\alpha'}{R}$ with the momenta and windings interchanged. When the compact background space is a torus $T^{n}$ characterised by some background moduli, T-duality implies that those backgrounds related by the non-compact group $O(n,n,\mathbb{Z})$ are physically equivalent. As we move to lower energies and hence go from string theory to the low energy effective supergravity, such exclusively stringy phenomena tend to be lost in the maze. Keeping track of them is usually not easy and a superficial approach may very well overlook certain aspects strictly related to stringy dualities. It has for example been shown that keeping explicit T-duality manifest by means of  double field theory may lead to moduli stabilisation and a substantial number of de-Sitter vacua solutions otherwise prohibited by a series of no-go theorems for geometric fluxes [2]. T-duality allows new (non-) geometries to be considered valid string backgrounds [3]. These appear as generalisations of standard Riemannian spaces and are usually called non-geometric string backgrounds. In order to describe the dynamics of a string in such a non-geometric background we need to consider the interplay between winding and momentum modes. Due to the non-geometric nature of the string background we may expect new phenomena, not present when only geometric backgrounds are considered. One particularly relevant example is a new kind of non-commutativity and non-associativity of the closed string coordinates in the presence of non-geometric fluxes [4] characterised by the relations 
\begin{equation}
[X^{I}(\tau,\sigma),X^{J}(\tau,\sigma)]\cong P^{K}
\end{equation}
respectively 
\begin{equation}
[[X^{I}(\tau,\sigma), X^{J}(\tau,\sigma), X^{K}(\tau,\sigma)]]\neq 0
\end{equation}
Like Heisenberg's uncertainty relations, these relations describe the limited resolution of a string's position [4]. Such effects arise at the interface between small and large compact dimensions. Standard quantum entanglement represents the impossibility of separating the description of entangled subsystems in terms of independent descriptions of the individual subsystems alone. Such impossibility arises from the standard commutation relations that quantum observables usually satisfy. String theoretical entanglement has however not been explored. When we consider the Neveu-Schwarz (NS/NS) sector which consists of the symmetric metric $g_{ij}$, the anti-symmetric $B$-field $B_{ij}$ and the dilaton $\phi$ as massless string excitations, we do not describe their full dynamics by means of string field theory, a task too difficult for now. Instead we consider a low energy limit and write an effective field theory given by 
\begin{equation}
S_{NS}=\int d^{D}x\sqrt{-g}e^{-2\phi}[\mathcal{R}+4(\partial \phi)^{2}-\frac{1}{12}H_{\mu\nu\rho}H^{\mu\nu\rho}]
\end{equation}
By construction, this action only considers strings and momentum modes. As a consequence, T-duality is not implemented and the theory is blind to non-geometric backgrounds depending on the interplay between winding and momentum modes. The fields $g_{ij}$, $B_{ij}$, and $\phi$ introduced here would therefore be ill defined when analysed in non-geometric (string) backgrounds. To make T-duality manifest at the level of effective field theories we employ the so called double field theory [5]. Once a working formulation of double field theory exists, an important question is to calculate the entanglement entropy associated to it. While for the cases in which $g_{ij}$, $B_{ij}$, and $\phi$ are well defined we may employ a geometric approach of the type defined by the Ryu-Takayanagi formula, in the non-geometric case, string-geometry effects become relevant. As entanglement entropy is usually UV divergent, it requires renormalisation. The counter-terms have been calculated for various special cases, yet, no general relation connecting them is known as for now. In what follows I will show that making T-duality manifest leads to new relations between entanglement entropy counter-terms allowing for a consistent renormalisation of entanglement entropy. Moreover, extending the geometry by taking into account the double field theory generalised metric will allow us to extend the geometric approach to entanglement entropy in contexts where usual geometric arguments would not easily apply. Corrections to the Ryu-Takayanagi formula of the type developed by Lewkowycz and Maldacena will be obtained by purely geometric (albeit non-Riemannian) considerations. 
In the next section I will present the basics of double field theory as well as the prescriptions used to generalise the metric and Lie derivatives. 

In section three I will present some remarks on the entanglement entropy and its holographic calculation as well as the basic idea behind the higher order corrections to the Ryu-Takayanagi formula. 

In section four, I will introduce the T-dual formulation of the AdS space following ref. [26] and I will analyse the effects of introducing the double field theory bulk on the construction of the entanglement entropy. In order to do this I will rely on a group manifold formulation of the double field theory. The manifest T-duality will be introduced for the AdS space and it will be noticed that the T-dual complement of AdS is actually the de Sitter space. I will also explain how the extended metric associated to the doubled AdS space can be used in the construction of the counter-terms meant to renormalise the entanglement entropy. Geometric arguments will be added to make the discussion more suitable for a holographic interpretation. 

In section five, I will connect double field theory with the holographic approach to the calculation of the entanglement entropy and I will present a systematic way of performing entanglement entropy renormalisation based on the manifest use of T-duality within double field theory. 

In section six, I will explain the effects of double field theory on the calculation of the entanglement entropy by means of minimal surfaces in the bulk spacetime.

In section seven, I will explain the way in which connections on a manifold are changed by the use of non-Riemannian geometry induced by double field theory and how curvatures are being computed. 

In section eight, I explain the basics of how the minimal surfaces emerge in the context of double field theory and how light sheets are being re-interpreted by using the Raychaudhuri equation [31].

In section nine I introduce the concept of extremal area both for de-Sitter and anti de-Sitter spacetimes as they arise in a double field interpretation.
Finally, in section ten I conclude by showing the similarity between Slavnov Taylor identities connecting counter-terms and insuring gauge invariance of the renormalisation approach and the relations between counter terms resulting from the preservation of explicit T-duality in the context of the renormalised entanglement entropy. 

\section{2. Double field theory}
Double field theory is an effective theory emerging from closed string theory which distinguishes itself from other effective field theories by making T-duality explicitly manifest [6]. T-duality is a fascinating symmetry of string theory which, among others, can relate string theories on different topologies. By doing this it also gives a proper formulation to the principle of topological invariance employed in [1] and  [37-39]. Double field theories are in general constrained theories. As the geometrization of T-duality implies generalising diffeomorphisms acting on the extended space, we obtain a new, extended diffeomorphism algebra which must be closed [7]. For this to happen various types of constraints can be introduced. The most restrictive one, that eliminates half of the degrees of freedom in double field theory, reducing it to supergravity, is the so called section constraint (or strong constraint). The strong constraint prohibits winding and momentum excitations in the same direction. Although this constraint abandons the original scope of double field theory, it is still capable of elucidating certain stringy features leading for instance to string corrections of the order $\alpha'$ [8]. If we do not wish to give up entirely on the double fields, there exist alternative constraints (known as Scherk-Schwarz type constraints [9]) capable of restoring algebra closure at a lower conceptual cost. These constraints appear in the Scherk-Schwarz compactification [10] which contain generic gaugings of gauged supergravity theories. The double coordinates are incorporated through the twist matrix and constant gaugings are computed by using this twist matrix [11]. Closure of the algebra is guaranteed if the above mentioned gaugings satisfy certain quadratic constraints [11]. In this case there is no need for a strong constraint. Relaxing the strong constraint allows us to search for truly double solutions of the equations of motion. Transcending supergravity in this way would allow us to find solutions that do not have any local interpretation from a supergravity point of view. Double field theory in general adds to the $D$ spacetime coordinates $x$ conjugate to the momentum modes, another $D$ coordinates $\tilde{x}$ conjugate to the winding modes of the string. These $2D$ coordinates are being combined into a $2D$-dimensional vector space with vectors described by $X^{M}=(\tilde{x}_{i}, x^{i})$. The partial derivatives with respect to the winding resp. momentum coordinates can also be combined into a vector $\partial_{M}=(\tilde{\partial}^{i}, \partial_{i})$. We define the group of $2D\times 2D$ matrices $h$ satisfying the condition $h^{t}\eta h=\eta$ and $h^{-1}=\eta^{-1}h^{t}\eta$ as $O(D,D)$. The associated Lie algebra has generators $T$ satisfying the relation $T^{t}\eta+\eta T=0$. They have the form 
\begin{equation}
   T=
  \left( {\begin{array}{cc}
   \alpha & \beta \\
   \gamma & \delta \\
  \end{array} } \right)
\end{equation}
where $\gamma$ and $\beta$ are antisymmetric and $\delta=-\alpha^{t}$. 
To rise or lower the index of the doubled vector we use the $O(D,D)$ invariant metric
\begin{equation}
   \eta^{MN}=
  \left( {\begin{array}{cc}
   0 & \delta_{i}^{j} \\
   \delta^{i}_{j} & 0 \\
  \end{array} } \right)
\end{equation}
Using the metric $g_{ij}$ and the anti-symmetric tensor $B_{ij}$ with $i,j=1,2,...,D$ we can construct a $2D\times 2D$ matrix with the property that it transforms as a tensor by a transformation from the $O(D,D)$ group. Indeed such a matrix will have the form 
\begin{equation}
\mathcal{H}_{MN}(X)=\left( {\begin{array}{cc}
   g^{ij} & -g^{ik}B_{kj} \\
   B_{ik}g^{kj} & g_{ij}-B_{ik}g^{kl}B_{lj} \\
  \end{array} } \right)
\end{equation}
The doubled space has therefore two metrics, one given by $\eta_{MN}$ used to lower and rise indices, and the other defined by $\mathcal{H}_{MN}$ which contains the dynamical fields. The double field effective action is then written as 
\begin{widetext}
\begin{equation}
S=\int dx \cdot d\tilde{x}\cdot e^{-2\phi}(\frac{1}{8}\mathcal{H}^{MN}\partial_{M}\mathcal{H}^{KL}\partial_{N}\mathcal{H}_{KL}-\frac{1}{2}\mathcal{H}^{MN}\partial_{N}\mathcal{H}^{KL}\partial_{L}\mathcal{H}_{MK}-2\partial_{M}\phi\partial_{N}\mathcal{H}^{MN}+4\mathcal{H}^{MN}\partial_{M}\phi\partial_{N}\phi)
\end{equation}
\end{widetext}
This is an action for the $g$, $B$, and $\phi$ fields where $g$ and $B$ are implicitly included by means of the generalised metric. 
The strong constraint can be written as 
\begin{equation}
\eta^{MN}\partial_{M}\partial_{N}...=0
\end{equation}
where the constraint operation above acts on arbitrary products of fields and gauge parameters. 
The action above is gauge invariant when the strong constraint is satisfied [12]. In double field theory we may define an extended gauge parameter $\xi^{M}=(\tilde{\xi}_{i},\xi^{i})$. The gauge transformation acting on the dilaton field is 
\begin{equation}
\delta\phi=-\frac{1}{2}\partial_{M}\xi^{M}+\xi^{M}\partial_{M}\phi
\end{equation}
and the action of the gauge transformation on the generalised metric is 
\begin{widetext}
\begin{equation}
\delta\mathcal{H}^{MN}=\xi^{P}\partial_{P}\mathcal{H}^{MN}+(\partial^{M}\xi_{P}-\partial_{P}\xi^{M})\mathcal{H}^{PN}+(\partial^{N}\xi_{P}-\partial_{P}\xi^{N})\mathcal{H}^{MP}
\end{equation}
\end{widetext}
This gauge transformation has been identified with a diffeomorphism and a generalised Lie derivative has been introduced as 
\begin{equation}
\hat{\mathcal{L}}_{\xi}\mathcal{H}^{MN}=\delta \mathcal{H}^{MN}
\end{equation}
The generalised Lie derivatives of the Kronecker tensor and of $\eta_{MN}$ vanish along arbitrary vector fields. As a result, the constraint that $\mathcal{H}$ is an $O(D,D)$ matrix is compatible with its gauge symmetry. 
The action is fixed by the constraint that it is invariant under general diffeomorphisms. When we parameterise them by $\xi$ we can write their action on a generalised vector $V^{M}$ as 
\begin{equation}
\delta_{\xi}V^{M}=\mathcal{L}_{\xi}V^{M}=\xi^{N}\partial_{N}V^{M}-V^{N}\partial_{N}\xi^{M}+\partial^{M}\xi_{N}V^{N}
\end{equation}
The generalised dilaton $e^{-2\phi}$ behaves like a scalar being hence a good generalised measure for integration. The closure condition for generalised diffeomorphisms [13] is
\begin{equation}
\mathcal{L}_{\xi_{1}}\mathcal{L}_{\xi_{2}}-\mathcal{L}_{\xi_{2}}\mathcal{L}_{\xi_{1}}=\mathcal{L}_{[\xi_{1},\xi_{2}]_{C}}
\end{equation}
where the generalised Lie bracket is 
\begin{equation}
[\xi_{1},\xi_{2}]_{C}=\frac{1}{2}(\mathcal{L}_{\xi_{1}}\xi_{2}-\mathcal{L}_{\xi_{2}}\xi_{1})
\end{equation}
The section constraint implies (at least locally) that all fields and gauge parameters depend only on one half of the coordinates in the doubled space. This constraint, namely the choice of a section, will break the standard $O(D,D)$ invariance. In order to keep T-duality manifest, after the choice of the section, we need to preserve a set of isometries corresponding to different duality frames [14]. This particular requirement gives rise to constraints both on the quantum corrections to the entanglement entropy and on the counter-terms in analogy to the Slavnov-Taylor identities in Yang-Mills theories. 

The analogue of the scalar curvature is the dilaton equation of motion. We have
\begin{widetext}
\begin{equation}
\begin{array}{c}
\mathcal{R}=4\mathcal{H}^{MN}\partial_{M}\partial_{N}\phi-\partial_{M}\partial_{N}\mathcal{H}^{MN}-4\mathcal{H}^{MN}\partial_{M}\phi\partial_{N}\phi+4\partial_{M}\mathcal{H}^{MN}\partial_{N}\phi\\
\\
+\frac{1}{8}\mathcal{H}^{MN}\partial_{M}\mathcal{H}^{KL}\partial_{N}\mathcal{H}_{KL}-\frac{1}{2}\mathcal{H}^{MN}\partial_{M}\mathcal{H}^{KL}\partial_{K}\mathcal{H}_{NL}\\
\end{array}
\end{equation}
\end{widetext}
and the action becomes 
\begin{equation}
S=\int dx\cdot d\tilde{x}\cdot e^{-2\phi}\cdot \mathcal{R}
\end{equation}
It has been shown that $\mathcal{R}$ defined in this way is a gauge scalar [15]. Moreover, the variation of $S$ with respect to the generalised metric $\mathcal{H}_{MN}$ gives rise to the $O(D,D)$ tensor $\mathcal{R}_{MN}(\mathcal{H},\phi)$ which appears to be a generalised Ricci tensor. We can already see that we have left the context of Riemannian geometry as the contraction of $\mathcal{R}_{MN}$ does not yield $\mathcal{R}$. It has also been shown in [16] that when we demand that these quantities are completely determined in terms of the physical fields, their values become identical zero. This is a particularity of double field theory which will have an impact on understanding the relations between entanglement entropy counter-terms. 
Up to this point a geometric description of double field theory has been given and consistent generalisations of the metric, the gauge transformations, and the Lie derivative in the context of double field theory have been presented. It remains to be seen how entanglement entropy can be derived from purely geometric considerations and thereafter, how can a double field theory approach help in understanding the corrections to the Ryu-Takayanagi entropy formula [17].

\section{3. Geometry and Entanglement entropy}
Given a pure quantum state, when analysed by an observer who has access only to a part of the system, such a state may appear mixed. having the reduced density matrix $\rho_{A}=Tr_{B}(\rho_{tot})$ the entanglement entropy is defined as the von-Neumann entropy of the reduced density matrix
\begin{equation}
S=-Tr_{A}(\rho_{A}\cdot \log \rho_{A})
\end{equation}
Entanglement entropy is a non-local, non-observable, UV-cutoff dependent quantity. It is of major importance in understanding various properties of quantum field theory. By being non-local it gives us access to quantum entanglement phenomena as well as to a better understanding of dualities linking different topologies. There exists a geometric formula for computing entanglement entropy both for static and dynamic spacetime geometries. If one divides a system into two parts $A$ and $B$, a key property of the entanglement entropy is that $S_{A}+S_{B}\geq S_{A\cup B}$. For quantum field theory defined over a spacetime, given its ground state, we can define two regions as above and analyse the entanglement between them. Entanglement occurs mostly at the boundary between the two regions and is proportional to the number of degrees of freedom at that boundary. Given a region $A$ we therefore expect 
\begin{equation}
S_{A}\sim \frac{\mathcal{A}(\partial A)}{a^{2}}+...
\end{equation}
where $\partial A$ is the boundary of the region $A$, $\mathcal{A}$ denotes the area of the argument, and $a$ is the spacing of the lattice which regularises the quantum field theory in this example. Thinking in holographic terms, one arrives at the conclusion that entropy in quantum gravity must satisfy an area law [18]. As a result, entanglement entropy of any spatial region $A$ in a holographic boundary can be expressed as the area of an extremal surface in the bulk spacetime
\begin{equation}
S_{A}=\min\limits_{X\sim A}\frac{\mathcal{A}(X)}{4G_{N}}
\end{equation}
 When analysing the bulk spacetime at low energies one uses Einstein's gravity corrected by higher derivative interactions. These are responsible for higher derivative corrections to the Ryu-Takayanagi formula [19]
 \begin{equation}
 S_{gen}=S_{Wald}+S_{extrinsic}
 \end{equation}
 the first term being the Wald entropy and the second being given by the corrections from the extrinsic curvature of the bulk minimal surface. $S_{gen}$ is the full classical gravitational entropy also called the generalised area. String theory predicts $\alpha'$-corrections to the entropy and Ryu-Takayanagi prescription can be extended to general theories of higher derivative gravity involving such corrections [20]. The formula proposed by Wald [21] is 
\begin{equation}
S_{Wald}=-2\pi\int d^{d}y\sqrt{g}\frac{\delta \mathcal{L}}{\delta R_{\mu\rho\nu\sigma}}\epsilon_{\mu\rho}\epsilon_{\nu\sigma}
\end{equation}
The terms involving the extrinsic curvature vanish at a Killing horizon but their differences matter in entanglement entropy calculations because minimal surfaces may have nonzero extrinsic curvature [19]. The extrinsic curvature terms do not appear in Einstein gravity. As showed in [19], higher order contributions to the action may be enhanced due to would-be logarithmic divergences. The extrinsic curvature corrections can be regarded as anomalies in the variation of the action. The idea of the Ryu-Takayanagi prescription is to use the replica trick within the bulk. The Renyi entropy is defined by 
\begin{equation}
S_{Renyi}=-\frac{1}{n-1}\log Tr[\rho^{n}]
\end{equation}
with $\rho$ the reduced density matrix of the subsystem $A$. Analytic continuation of the Renyi entropy for $n\rightarrow 1$ leads to the entanglement entropy. But at the same time, the Renyi entropy can be calculated using the partition function of the field theory on an $n$-fold cover
\begin{equation}
S_{Renyi}=-\frac{1}{n-1}(\log Z_{n}-n\log Z_{1}). 
\end{equation}
The initial spacetime manifold $M_{1}$ is extended to $M_{n}$ by taking $n$ copies of itself and cutting each at the spatial region of interest for our theory, namely $A$. We then glue them all together in a cyclic order. When a holographic dual is available, the bulk solution for the extended manifold is $B_{n}$ having $M_{n}$ as boundary. We can then identify the field theory partition function on the extended boundary $M_{n}$ with the bulk action in the large $N$ limit [19]
\begin{equation}
Z_{n}=Z[M_{n}]=e^{-S[B_{n}]+...}
\end{equation}
The extension of $M_{n}$ to non-integer $n$ is not generally defined. On the bulk side however it is possible to analytically continue an orbifold of $B_{n}$ to non-integer $n$. Note that the boundary $M_{n}$ at integer $n$ has a $Z_{n}$ symmetry. This translates in a cyclic permutation symmetry of the $n$ replicas. If this symmetry extends to the bulk we may consider the bulk orbifold $\hat{B}_{n}/Z_{n}$ which has singularities at the fixed points of the $Z_{n}$ action. The fixed points are described by a co-dimension $S$ surface presenting a conical defect. This surface ends on the boundary of our region of interest $\partial A$. The boundary of $\hat{B}_{n}$ is the original manifold $M_{1}$. Let the angle around the boundary $\partial A$ be $\tau$, then the $Z_{n}$ symmetry acting on the boundary $B_{n}$ translates into $\tau\rightarrow\tau+2\pi$. The set of fixed points forms a co-dimension 2 surface which ends on $\partial A$. This surface will be called $C_{n}$. For an integer $n$, the bulk solution $B_{n}$ must be regular in the interior, hence its $Z_{n}$ orbifold has a conical defect at $C_{n}$ with opening angle $\frac{2\pi}{n}$. But we know that the entropy can be written as $S[B_{n}]=n\cdot S[\hat{B}_{n}]$ at integer $n$. Here $S[\hat{B}_{n}]$ does not include any contribution from the conical defect. At the asymptotic boundary $M_{1}$ we need to include in $S[\hat{B}_{n}]$ the usual Gibbons-Hawking-York surface term as well as counterterms. By analytic continuation of $\hat{B}_{n}$ towards non-integer $n$ we obtain $S[B_{n}]$. Using the Renyi entropy formula and the gauge gravity duality formula we obtain 
\begin{equation}
S_{n}=\frac{n}{n-1}(S[\hat{B}_{n}]-S[\hat{B}_{1}])
\end{equation}
Expanding around $n=1$ gives us the entanglement entropy. In the $n\rightarrow 1$ limit, $C_{n}$ approaches the minimal surface of the Ryu-Takayanagi formula. Considering $\rho$ as parametrising the minimal distance to $C_{n}$ the metric becomes 
\begin{equation}
ds^{2}=\rho^{-2\epsilon}(d\rho^{2}+\rho^{2}d\tau^{2})+(g_{ij}+2K_{aij}x^{a})dy^{i}dy^{j}+...
\end{equation}
where the indexes $a,b,...$ refer to the coordinates in the $(\rho,\tau)$ plane orthogonal to $C_{n}$ while the $i,j,...$ indexes refer to the coordinates along $C_{n}$. The term $K_{aij}$ represents the extrinsic curvature tensor of the co-dimension $2$ surface $C_{n}$. The angular coordinate around the area of interest $A$ has the range $\tau \in [0,2\pi)$ and the conical deficit of this metric at $\rho=0$ is $2\pi\epsilon$. The deficit at integer $n$ is $2\pi-2\pi/n$. For these two to agree we set $\epsilon=1-\frac{1}{n}$. We can re-write the above metric in a regularised form as
\begin{equation}
ds^{2}=e^{2A}(d\rho^{2}+\rho^{2}d\tau^{2})+(g_{ij}+2K_{aij}x^{a})dy^{i}dy^{j}+...
\end{equation}
where the regulator $A$ contains the thickness parameter $a$ in the formula
\begin{equation}
A=-\frac{\epsilon}{2}\log (\rho^{2}+a^{2})
\end{equation}
The final result does not depend on the regulator as the coefficient of the logarithmic divergence is universal. The terms contributing linearly in $\epsilon$ are the Wald term and the anomaly term. The Wald term originates in a single Riemann tensor 
\begin{equation}
R_{z\bar{z}z\bar{z}}=\frac{1}{4}e^{2A}\hat{\nabla}^{2}A+...
\end{equation}
where $\hat{\nabla}$ means covariant derivative with respect to the metric with the singular warp factor eliminated i.e. $\hat{g}_{ab}=e^{-2A}G_{ab}$. Performing the calculations according to [19] we determine the contribution to the entanglement entropy as predicted by Wald. 
The anomaly term originates from the product of two Riemann tensors, one of the form 
\begin{equation}
R_{zizj}=2K_{zij}\nabla_{z}A+... 
\end{equation}
while the other being the conjugate in two indexes $R_{\bar{z}k\bar{z}l}$. 
Therefore the contribution to the regularised cone to be considered now is 

\begin{equation}
S^{(2)}=4^{2}\int d^{D}x\sqrt{G}\frac{\delta^{2} \mathcal{L}}{\delta R_{zizj}\delta R_{\bar{z}k\bar{z}l}}(2K_{zij}\nabla_{z}A)(2K_{\bar{z}ij}\nabla_{\bar{z}}A)
\end{equation}
The factor $4^{2}$ just keeps track of equivalent forms of the Riemann tensor. Following the calculation of the entanglement entropy in [19] we obtain the correction of the form 
\begin{equation}
S^{(2)}_{EE}=16\pi\int d^{d}y\sqrt{g}\sum_{\alpha}(\frac{\delta^{2}\mathcal{L}}{\delta R_{zizj}\delta R_{\bar{z}k\bar{z}l}})_{\alpha}\frac{K_{zij}K_{\bar{z}kl}}{q_{\alpha}+1}
\end{equation}
where $q_{\alpha}$ can be thought of as anomaly coefficients in the interpretation of [19]. In this sense, the extrinsic curvature corrections to the Wald formula, as noted above, can be seen as anomalies. Indeed in the case of Weyl symmetry, we had a regulated effective action $W$ which splits into a renormalised effective action $W_{fin}$ and a counter-term $W_{ct}$. As the full action $W$ is invariant under a Weyl transformation and a compensating scaling of the cut-off, the variation of $W_{fin}$ under such a transformation must therefore be compensated by a equal variation with opposing sign of $W_{ct}$. The same happens for our entanglement entropy in the present case. 

\section{4. T-dual anti-de-Sitter space and the covariant derivative}
Interpreting the holographic entanglement entropy from a string theoretical perspective is a particularly difficult task as it involves several aspects of quantum gravity in the bulk for which we do not have the required computational tools. Nevertheless, one particular stringy feature, namely T-duality, allows us to perform certain computations taking fundamentally stringy phenomena into account. Several questions arise when we try to understand entanglement from a geometric perspective while including string theoretical effects. First, the best understood holographic duality relates anti-de-Sitter spaces in the bulk with conformal field theories on the boundary. Analysing the bulk is usually difficult. Analysing the bulk in the context of manifest T-duality, and hence including string theoretical effects is particularly difficult due to various non-geometric aspects related to manifolds with explicit T-duality. Therefore one first task will be to formulate AdS space with manifest T-duality in a form reminding us of double field theory. This has fortunately been analysed in [26] as well as in [27], [28], and [29]. The group theoretical approach of reference [26] is most suited for the present discussion and will be used as an introduction to the subject. But even when we do have a well defined doubled AdS space with manifest T-duality, we do need a way to constrain the resulting theory such that we obtain again the physical results, while keeping the string features encoded by T-duality manifest. The use of the dimensional reduction constraint in ref. [26] is related to the section condition as will be shown at the proper time. Once the algebraic structure for the doubled AdS space is well defined we have to work on the geometry generated in this way. It is worthwhile to mention that the extended metric will be composed of three components as follows
\begin{equation}
\eta_{\underline{a}\, \underline{b}}=(\eta_{\natural\natural},\eta_{a\,b},\eta_{a'\, b'})
\end{equation}
where $\eta_{\natural\natural}$ represents the common direction, $\eta_{a\,b}$ represents the AdS space and $\eta_{a'\, b'}$ represents the T-dual. It will result that the T-dual space metric is that of the de-Sitter space, leading to additional terms that will have a role in the construction of the counter-terms for the entanglement entropy. The formulation of the covariant Ryu-Takayanagi formula in the doubled AdS space will require the proper definition of the covariant derivative and of the connection on the doubled space, considering that the procedure of doubling introduces stringy phenomena. 
Entanglement entropy is known to depend on geometry and to be divergent. Holographic entanglement entropy has been calculated up to now within the AdS/CFT duality resulting in the well known Ryu-Takayanagi formula 
\begin{equation}
S_{A}=\frac{\min(\mathcal{A}(\partial A))}{4G_{N}}
\end{equation}
where the numerator refers to the minimal area of the bulk surface within the AdS space bounded by $A$ in the boundary. It should be noted that the doubled space in the bulk will introduce additional terms, at least due to the extended metric arising in double field theory. This metric combines anti-de-Sitter and de-Sitter spaces leading to a dynamics where a separation between left modes and right modes becomes natural, with the left modes moving in the anti-de-Sitter space and the right modes moving in the de-Sitter space [26]. 

In fact, the manifestly T-dual formulation is a gauge theory that takes into account certain aspects of stringy gravity. As a gauge theory, it will have a gauge field, denoted following [26] as $A_{m}^{\;\;I}$ and a gauge group denoted as $G_{I}$. In this case the covariant derivative, the gauge transformation rule, and the field strength become
\begin{equation}
\begin{array}{c}
p_{m}=\frac{1}{i}\partial_{m}\rightarrow \nabla_{m}=p_{m}+A_{m}^{\;\;I}G_{I}\\
\\
\delta_{\lambda}\nabla_{m}=i[\Lambda,\nabla_{m}]\\
\\
\Lambda=\lambda^{I}G_{I}\Rightarrow \delta_{\lambda}A_{m}=-\partial_{m}\lambda\\
\\
i[\nabla_{m},\nabla_{n}]=F_{mn}^{\;\;\;\;I}G_{I}\\
\end{array}
\end{equation}
Gravity is a gauge theory of the vielbein $e_{a}^{\;\; m}$. In this case we obtain the covariant derivative and the general coordinate transformation rule as 
\begin{equation}
\begin{array}{c}
p_{m}\rightarrow \nabla_{a}=e_{a}^{\;\;m}p_{m}\\
\\
\delta_{\lambda}\nabla_{a}=i[\Lambda,\nabla_{a}]\\
\\
\Lambda=\lambda^{m}p_{m}\Rightarrow\\
\\
\Rightarrow \delta_{\lambda}e_{a}^{\;\;m}=\mathcal{L}_{\lambda}e_{a}^{\;\;m}=\lambda^{n}\partial_{n}e_{a}^{\;\; m}-e_{a}^{\;\;n}\partial_{n}\lambda^{m}\\
\end{array}
\end{equation}
Together with the Lorentz generator $s_{mn}$ we include the curvature tensor 
\begin{equation}
\begin{array}{c}
\nabla_{a}=e_{a}^{\;\;m}p_{m}+\frac{1}{2}\omega_{a}^{\;\;mn}s_{mn}\\
\\
i[\nabla_{a},\nabla_{b}]=T_{ab}^{\;\;\;\;c}\nabla_{c}+\frac{1}{2}R_{ab}^{\;\;\;\;cd}s_{cd}
\end{array}
\end{equation}
The torsion constraint $T_{ab}^{\;\;\;\;c}=0$ relates $e_{a}^{\;\;m}$ and $\omega_{a}^{\;\;mn}$. 
The T-duality effects of stringy gravity are included by doubling the momentum coordinates through the introduction of the winding modes, namely by performing the replacement 
\begin{equation}
p_{m}\rightarrow (p_{m}(\sigma),\partial_{\sigma}x^{m}(\sigma))
\end{equation}
The gauge theory for stringy gravity is associated to the doubled vielbein field $e_{\underline{a}}^{\;\;\underline{m}}\in O(d,d)/O(d-1,1)^{2}$. Using the notation of [26], the stringy covariant derivative would be the next step in our generalisation 
\begin{equation}
p_{m}\rightarrow \nabla_{a}\rightarrow \rhd_{\underline{a}}(\sigma)=e_{\underline{a}}^{\;\;m}p_{m}+e_{\underline{a}m}\partial_{\sigma}x^{m}
\end{equation}
and the associated gauge transformation rule becomes 
\begin{equation}
\begin{array}{c}
\delta_{\lambda}\rhd_{\underline{a}}(\sigma)=i[\Lambda,\rhd_{\underline{a}}(\sigma)]\\
\\
\Lambda=\int d\sigma (\lambda^{m}p_{m}+\lambda_{m}\partial_{\sigma}x^{m})
\end{array}
\end{equation}
Similarly, the curvature tensors are being obtained by introducing the Lorentz generator. We will have left and right Lorentz generators $S_{\underline{m}\;\underline{n}}=(S_{mn},S_{m'n'})$ defined respectively in their left and right bases $P_{\underline{m}}=(P_{m},P_{m'})$
\begin{equation}
\begin{array}{cc}
P_{m}=\frac{1}{\sqrt{2}}(p_{m}+\partial_{\sigma}x^{m}), & P_{m'}=\frac{1}{\sqrt{2}}(p_{m}-\partial_{\sigma}x^{m})
\end{array}
\end{equation}
in the unitary gauge. At this point we can go to the fully doubled vielbein $E_{\underline{A}}^{\;\;\;\underline{B}}$ obtaining the fully stringy covariant derivative 
\begin{equation}
\rhd_{\underline{A}}=E_{\underline{A}}^{\;\;\;\underline{M}}\overset{\circ}{\rhd}_{\underline{M}}
\end{equation}
The circled operators refer to the AdS space while the other operators refer to the operators associated to gravity coupled to the AdS space. Such gravity theories appear as fluctuations in the asymptotically AdS space. This covariant derivative defines the type of geometry to be considered when a string theoretical formulation of the Ryu-Takayanagi formula is derived. The metric in the case of doubled $AdS_{5}$ will be [26]
\begin{equation}
\eta_{\underline{a}\,\underline{b}}=(-1;-1,1,1,1,1;1,-1,-1,-1,-1)
\end{equation}
Using this metric, before imposing any restrictive constraints that would eliminate the additional coordinates, results in a symmetry in the geometric terms arising in the entanglement entropy. It can already be seen that by considering such stringy effects in the bulk, the entanglement entropy as calculated from geometric arguments will receive corrections that will appear in symmetric pairs that would imply the addition of counter-terms that would precisely cancel the divergent parts at all orders. In order to perform such a computation we must first understand this extended geometry in the bulk as well as its physical effects. As a criterium for the construction of the double AdS algebra we demand that the dimensional reduction gives us the original AdS algebra with the normal single coordinate space. Following ref. [26] we also assume that the doubled AdS algebra reduces in the large AdS radius limit to the flat space. Also, the doubled AdS algebra has a nondegenerate group metric and the structure constant is totally antisymmetric. We will work with group manifolds and we double the AdS group by introducing the left and the right AdS groups. There will be a left-right mixing to be taken into account. The associated algebra will be generated by the momenta $p_{\underline{a}}=(p_{a},p_{a'})$, and the doubled Lorentz generators $s_{\underline{a}\,\underline{b}}=(s_{ab}, s_{a'b'}; s_{ab'})$ with $a,b = 0,...,d-1$.
In the general case, the string covariant derivative $\rhd_{I}(\sigma)$ can be constructed using the $B$-field from the particle covariant derivative $\nabla_{I}(\sigma)$ and the $\sigma$ component of the left invariant current $J_{1}^{\;\;I}(\sigma)$
\begin{equation}
\rhd_{I}=\nabla_{I}+\frac{1}{2}J_{1}^{K}(\eta_{KI}+B_{KI})
\end{equation}
Generalising this for the AdS space we obtain a linear combination of the AdS particle covariant derivative $\overset{\circ}{\nabla}_{\underline{A}}$ and the $\sigma$ component of the left invariant current $\overset{\circ}{J^{\underline{A}}}$ with the $B$-field
\begin{equation}
\overset{\circ}{\rhd}_{\underline{A}}=\overset{\circ}{\nabla}_{\underline{A}}+\frac{1}{2}\overset{\circ}{J^{\underline{B}}}(\eta_{\underline{B}\,\underline{A}}+\overset{\circ}{B}_{\underline{B}\,\underline{A}})
\end{equation}
At this point we have the covariant derivative of the doubled AdS space which now allows us to calculate the entanglement entropy from geometric considerations taking into account the stringy nature of the bulk space.

\section{5. Holographic entanglement entropy in the bulk}
The connection between calculating entanglement entropy in a quantum field theory on the conformal boundary and calculating an extremal area in the bulk theory represents a practical realisation of the holographic principle.
 In order to understand this type of calculations in the simple setting of an AdS bulk space, let me start with a heuristic development of the covariant entropy bound as given by [30]. 

Already at the level of constructing the minimal surface in an asymptotically AdS spacetime, problems arise due to the fact that the Lorentzian spacetimes do not have a definite metric signature. This problem will become more acute in the T-dual case as the extended metric will contain a de-Sitter part with additional time-like directions. For the simplest case however, reference [30] offers a good argument in favour of a minimisation principle for areas within an AdS space only. I will re-state those and extend them for the T-dual case where de-Sitter spacetimes will arise as well. Consider the boundary theory as being in a time varying state on a fixed background taken to be the boundary $\partial \mathcal{M}$ of a bulk spacetime $\mathcal{M}$. The bulk geometry corresponding to such a time varying state will have an explicit time dependence despite the fact that the boundary background spacetime is fixed. Because of this, the bulk spacetime will not have a time-like Killing field. We can still choose a foliation by equal time slices on the non-dynamical boundary background metric $\partial\mathcal{M}$. This can be done in a way such that the time coordinate will implement the natural Hamiltonian evolution of the field theory $\partial\mathcal{M}=\partial\mathcal{N}_{t}\times \mathbb{R}_{t}$. Now take a region of the spacelike boundary part $\mathcal{A}_{t}\in \partial\mathcal{N}_{t}$. The calculation in terms of path integrals could proceed as known. We could also implement a minimal surface prescription, however, now we have to pay attention to the fact that the spacetime is Lorentzian and hence the equal time foliation on the boundary $\partial \mathcal{M}$ does not necessarily correspond to a natural foliation of the bulk $\mathcal{M}$. This problem will manifest itself in the case of the T-dual prescription as well. Once a natural foliation has been identified, one can compute the holographic entanglement entropy by using a preferred spacelike slice $\mathcal{N}_{t}\subset \mathcal{M}$ defined by the extension of the slice from the boundary $\partial\mathcal{M}$. The metric induced on $\mathcal{N}_{t}$ is spacelike and hence the minimal surface is well defined. Therefore, in general we wish to find a minimal surface $\mathcal{S}\in\mathcal{N}$ such that $\partial\mathcal{S}_{\partial\mathcal{M}}=\partial\mathcal{A}$. We look for a spacelike slice of the bulk which is covariantly well defined, which is anchored on $\partial\mathcal{N}_{t}$ and which reduces to the constant time slice for a static bulk. In general there is no preferred time slicing of $\mathcal{M}$. The AdS spacetime however has a natural foliation in terms of zero mean curvature slices. Each such slice corresponds to a maximal area spacelike slice through the bulk, attached at the boundary slice $\partial\mathcal{N}$. The leaves of this maximal area foliation are denoted by $\Sigma_{t}$. Unless one covers all spacelike directions, the maximal area slice is not well defined as, given a surface, one can increase the area for the same anchoring boundary by wiggling the surface in another spacelike direction. In [30] this problem has been avoided by assuming that the surface considered was co-dimension one, hence extending over all available spatial directions and leaving no direction for the disturbing deviations. Also, the area of any spacelike slice in AdS spacetimes is manifestly infinite. This can however be regulated such that the computationally relevant quantities become finite. On the slices $\Sigma_{t}$ we consider the minimal area surface anchored at $\partial\mathcal{A}_{t}$. The holographic entanglement entropy can be calculated by means of a mini-max algorithm. Firs one has to find the maximal slice in the bulk which agrees with the spacelike foliation of the boundary $\partial\mathcal{M}$ and once such a slice is found, one has to find within it a minimal surface named $\mathcal{X}$. The entanglement entropy is then given by the area of $\mathcal{X}$ as 
\begin{equation}
S_{\mathcal{A}}=\frac{Area(\mathcal{X})}{4G_{N}^{(d+1)}}
\end{equation}
A more natural minimal surface in the context of the holographic principle can be constructed by means of light-sheets given the covariant entropy bounds in gravitational theories.

Let us therefore focus on the light-sheets and derive the entropy bounds in a geometric context. The standard starting point is the context of the AdS/CFT duality with a $d+1$ dimensional AdS space $\mathcal{M}$ and its boundary $\partial\mathcal{M}$ of dimension $d$. At a given time $t_{0}$ we divide the $d-1$ dimensional spacelike boundary into two pieces $\mathcal{A}_{t_{0}}$ and $\mathcal{B}_{t_{0}}$. The boundary between these domains will be designated as $\partial\mathcal{A}_{t}$ and will be a spacelike surface of dimension $d-2$ in $\partial\mathcal{M}$. Using the conformally flat metric on $\partial\mathcal{M}$ we can construct the upper and lower light-sheets in the boundary space as $\partial L_{t_{0}}^{+}$ and $\partial L_{t_{0}}^{-}$. We can extend those light-sheets from the boundary towards the bulk and call the extensions $L_{t_{0}}^{\pm}$. They will become the light-sheets in the bulk space $\mathcal{M}$. We define the spacelike surface $\mathcal{Y}_{t_{0}}=L_{t_{0}}^{+}\cap L_{t_{0}}^{-}$. Following [30] the entanglement entropy is defined by 
\begin{equation}
S_{\mathcal{A}_{t_{0}}}(t)=\frac{\min_{\mathcal{Y}}(Area(\mathcal{Y}_{t_{0}}))}{4G_{N}^{(d+1)}}
\end{equation}
By $\min_{\mathcal{Y}}(Area(\mathcal{Y}))$ we understand taking the minimum over the set of possible values of $\mathcal{Y}$ as we vary the form of $L_{t_{0}}^{\pm}$ given the fixed boundary $\partial L_{t_{0}}^{\pm}$. Clearly in order to define light-sheets we need to have a spacelike surface in our manifold and we construct the four congruences of future/past null geodesics from that surface towards the in-going and out-going directions. The light-sheet for that surface corresponds then to those null geodesic congruences for which the expansion of the null geodesics is non-positive definite. Physically, the requirement is that the cross-sectional area at a given constant affine parameter along the congruence does not increase [30].
This is a valid request when thinking in terms of AdS spacetime, however, by T-duality, we are going to obtain the T-dual of the AdS spacetime which is a de-Sitter spacetime. In this case the criterium of non-positivity of the expansion is not maintained. The combination of AdS spacetime and its T-dual will impact the renormalisation of entanglement entropy due to the terms introduced by the de-Sitter spacetime. Let us first discuss the AdS case. 
It is clearly important to calculate the expansion of the null geodesics given the bulk space, as this quantity is defining for the light-sheets and for the entire subsequent construction. Considering a spacetime manifold and a co-dimension two surface $\mathcal{S}$ defined by two constraints 
\begin{equation}
\begin{array}{cc}
\phi_{1}(x^{\nu})=0, & \phi_{2}(x^{\nu})=0\\
\end{array}
\end{equation}
there exist two one-forms $\nabla_{\nu}\phi_{i}(x^{\nu})$, $i=1,2$. By the requirement of non-degeneracy, the one-forms are linearly independent and hence we have the null form 
\begin{equation}
\nabla_{\nu}\phi_{1}+\mu\nabla_{\nu}\phi_{2}
\end{equation}
for two distinct values of $\mu$. In this way, two null-vectors have been constructed [30] having the form 
\begin{equation}
N_{\pm}^{\mu}=g_{\mu\nu}(\nabla_{\nu}\phi+\mu_{\pm}\nabla_{\nu}\phi_{2})
\end{equation}
which can be normalised by means of the relation 
\begin{equation}
N_{+}^{\mu}N_{-}^{\nu}g_{\mu\nu}=-1
\end{equation}
The overall bulk metric induces a metric on the surface $\mathcal{S}$ which we will name $h_{\mu\nu}$. The null extrinsic curvatures of this surface can be written in terms of our null vectors and this induced metric as 
\begin{equation}
(\chi_{\pm})_{\mu\nu}=h^{\rho}_{\;\;\mu}h^{\lambda}_{\;\;\nu}\nabla_{\rho}(N_{\pm})_{\lambda}
\end{equation}
An orthogonal null geodesic congruence expanded to the surface can be calculated as the trace of the null extrinsic curvature as in [30]
\begin{equation}
\theta_{\pm}=(\chi_{\pm})^{\mu}_{\;\;\mu}
\end{equation}
The null expansions represent a measure for the rates of change of the area of the surface $\mathcal{S}$ propagated along the null vectors. The map $X^{\mu}(\xi^{\alpha}):\mathcal{S}\rightarrow \mathcal{M}$ represents the embedding of our surface in the bulk spacetime. $\xi^{\alpha}$ represents a set of coordinates on the surface $\mathcal{S}$. Given an infinitesimal deformation $\delta X^{\xi^{\alpha}}$ orthogonal to $\mathcal{S}$ with fixed boundary conditions, the change in the area of the surface is given by 
\begin{equation}
\delta A\propto\int_{\mathcal{S}}(\theta_{+}N_{+}^{\mu}\delta X_{\mu}+\theta_{-}N_{-}^{\mu}\delta X_{\mu})
\end{equation}
The proportionality constant is positive. The identification of the sign of the null expansions $\theta_{\pm}$ leads to the identification of the sign of the variation of the area at an infinitesimal deformation. The surfaces with vanishing null expansions are clearly extremal surfaces i.e. saddle points of the area functional.

Starting with a standard $AdS_{d+1}$ metric 
\begin{equation}
ds^{2}=\frac{d\rho^{2}}{4\rho^{2}}+\frac{1}{\rho}\eta_{\mu\nu}dx^{\mu}dx^{\nu}
\end{equation}
where the conformal boundary is reached when $\rho\rightarrow 0$ and $\eta_{\mu\nu}$ is the Minkovski metric the entanglement entropy formula following Ryu-Takayanagi for a co-dimension two surface in the bulk is 
\begin{equation}
S=\frac{1}{4G_{d+1}}\int_{\Sigma}d^{D}\sigma^{\alpha}\sqrt{\gamma}
\end{equation}
$G_{d+1}$ is the gravitational constant and $\gamma$ is the determinant of the metric induced on the surface inside the bulk. In the limit $\rho\rightarrow 0$ there appear area divergences which can be regulated by a cutoff $\rho=\epsilon$. The renormalised entanglement entropy has been heuristically formulated as 
\begin{equation}
S_{cl-ren}=\mathcal{L}_{\epsilon \rightarrow 0}(S_{reg}-S_{ct})
\end{equation}
where the counter-term $S_{ct}$ represents a set of terms depending on the boundary of the minimal surface and the cutoff surface. The cutoff surface can be described in terms of an induced metric $h_{\mu\nu}$. The metric on the boundary of the minimal surface, $\tilde{\gamma}_{ab}$ is the extension of the minimal surface metric $\gamma_{ab}$. The Ricci scalar of the boundary of the minimal surface is $\mathcal{R}$. The associated Ricci tensor is $\mathcal{R}_{ab}$. The extrinsic curvature of the minimal surface from the perspective of the cutoff surface is $\mathcal{K}_{ab}$ with the trace $\mathcal{K}$. Following [23] the counter-terms must be functionals of the extrinsic and intrinsic curvature invariants and therefore 
\begin{equation}
S_{ct}=\int_{\partial \Sigma}d^{D-1}\sqrt{\tilde{\gamma}}\mathcal{L}(\mathcal{K},\mathcal{R},\mathcal{R}_{ab}\mathcal{R}^{ab},\mathcal{K}_{ab}\mathcal{K}^{ab},...)
\end{equation}

\section{6. Entanglement entropy with double field theory in the bulk}

Entanglement entropy is expected to be T-duality invariant. This has been verified in all cases computed up to now and results from the fact that it is a measure of the number of microscopic degrees of freedom which should not depend on duality transformations. One may verify the invariance of entanglement entropy to T-duality in a semi-classical context by checking that Buscher transformations do not change it. The analysis of the T-duality invariance of entropy in the context of double field theory has also been done in [14]. In order to preserve T-duality after the choice of a section or after imposing the constraint that the generalised Lie algebra closes, we have to preserve certain isometries in the doubled space. In double field theory, isometries are defined by setting the generalised Lie derivative of the generalised metric to zero [4]
\begin{equation}
\begin{array}{cc}
\mathcal{L}_{K_{I}^{\;J}}\mathcal{H}^{MN}=0, & \mathcal{L}_{K_{I}^{\;J}}\phi'=0
\end{array}
\end{equation}
where $K_{I}^{\;J}$ are the Killing vectors associated to the isometries. 
As mentioned in [4], the isometries give rise to homogeneous doubled spaces with a constant generalised Ricci scalar. 
 It has been shown that divergences of the entanglement entropy can be resolved by means of a special type of renormalisation. Also quantum corrections due to the entanglement between the bulk region encompassed by our minimal area and the rest of the bulk can be expressed meaningfully. However, there has been no attempt to reconcile these results with the requirement of T-duality invariance of entropy. The definition of entropy in a manifestly T-duality invariant context has been done in [14]. The calculation of entanglement entropy up to one loop in quantum corrections has been done in [22]. The resulting expression there was 
\begin{widetext}
\begin{equation}
S(A)=S_{cl}(A)+S_{bulk-ent}+\frac{\delta A}{4G_{N}}+<\Delta S_{W-like}>+S_{counter}+O(G_{N})
\end{equation}
\end{widetext}
where $S(A)$ is the full entanglement entropy, $S_{cl}(A)$ is the standard Ryu-Takayanagi formula $S_{cl}(A)=\frac{min(A)}{4G_{N}}$ which itself may be corrected by higher derivative gravity and may present area type divergencies as noted by [23], $S_{bulk-ent}$ is the entanglement within the bulk between the inner minimal surface region and the rest of the bulk calculated at one loop level, $\frac{\delta A}{4G_{N}}$ represents the change in the area due to the shift in the classical background arising because of quantum corrections, $<\Delta S_{W-like}>$ represents the Wald-like entropy term, $S_{counter}$ represent the counter-term corrections required to make the calculation finite. 
All the terms in 
\begin{widetext}
\begin{equation}
S_{quantum}=S_{bulk-ent}+\frac{\delta A}{4G_{N}}+<\Delta S_{W-like}>+S_{counter}+O(G_{N})
\end{equation}
\end{widetext}
refer to the higher order quantum corrections, including the counter-terms required to render the corrections finite. Counter-terms are required in the semi-classical part too in order to renormalise area divergences, as done in [23]. 
The renormalisation of entanglement entropy has been performed in [23] for the case of an anti-de-Sitter bulk without explicit use of T-duality. 

Several fundamentally new features arise when we try to explicitly introduce T-duality for the bulk space.

As has been noted in [22], the quantum corrections are given by the bulk entanglement entropy, particularly by the entanglement between the minimal bulk region and the rest of the bulk. Therefore, we could think of the bulk theory as an effective field theory living on a fixed background geometry and then analyse how the minimal surface theory would be entangled with the rest of the bulk in the same way as for any quantum field theory [22]. One may notice however the recursive nature of this procedure. At each order in the series of quantum corrections we identify another effective theory to be the boundary structure of yet the next order bulk theory and so on. In standard quantum field theory, gauge symmetry imposes a set of identities on these corrections known as Slavnov-Taylor identities (in the case of non-abelian gauge theories). This article is motivated by the suspicion that similar relations may exist between higher order quantum entanglement entropy corrections and between the counter-terms needed to render the result finite, with the underlying symmetry being T-duality, imposed such that the overall result for the entanglement entropy does not change when T-duality transformations are performed. To push the analogy further, in the same way in which the requirement of gauge invariance establishes a set of relations between the (unobservable) Green functions in a quantum field theory, the requirement of T-duality invariance establishes a set of relations that must be obeyed by quantum corrections and counter-terms in the entanglement entropy. In double field theory the section (or strong constraint) explicitly eliminates the field dependence on the double field coordinates and hence reduces de-facto the double field theory to standard supergravity. However, the strong constraint has not been solved in general and hence additional string-theoretical features may still hide in less trivial solutions of this constraint. For example it has been shown in [16] that in spite of the strong constraint, the constrained double field theory inherits from string theory a left-right structure not present in the usual quantum field theory. Moreover, weaker constraints may still generate consistent double field theories while keeping the double field coordinate dependence manifest.
T-duality must therefore take into account two types of counter-terms, those resulting from the renormalisation of the area divergencies and those resulting from preserving the finiteness of the higher order quantum corrections as done in [22]. Double field theory involves both these domains and provides relations between the two types of counter-terms. This observation is justified by the fact that the replica trick prescription for quantum corrections to entanglement entropy implies the calculation of the partition function of all bulk quantum fields and hence of the associated functional determinants for the fluctuations around the bulk geometries. T-duality however connects manifestly distinct scales and hence imposing it in the context of double field theory will result in isometries that will be defined over the whole bulk space, therefore playing the role of a common ground on which both the area counter-terms and the quantum correction counter terms are defined. Following the computation of [22], let $\tau$ be the time direction. The quantum partition function in the replica-trick is
\begin{equation}
\begin{array}{c}
Z_{q,n}=Tr[P \exp(-\int_{0}^{2\pi n}d\tau H_{b,n}(\tau))]=Tr[\hat{\rho}_{n}^{n}]\\
\\
\hat{\rho}=P \exp(-\int_{0}^{2\pi}H_{b,n}(\tau))\\
\end{array}
\end{equation}
$H_{b,n}(\tau)$ is a local integral over a constant $\tau$ spatial slice and can be viewed as the time dependent hamiltonian propagating the system in the time $\tau$. The density matrix $\hat{\rho}$ is a bulk quantity associated to the bulk geometry. 
When going to the doubled space, T-duality is manifest in the form of isometries of the generalised metric. However, the geometry of double field theory is based on generalised diffeomorphisms. To define a generalised connection on such a space one looks at the double space diffeomorphism covariant derivative [24]
\begin{equation}
\mathcal{D}_{\mu}=\partial_{\mu}-\mathcal{L}_{A_{\mu}}
\end{equation}
The natural connection compatible with both the generalised metric and the $O(D,D)$ structure and allowing the integration measure to contain the dilaton term $e^{-2\phi}$ is not uniquely defined. It has been shown in [25] that the connection coefficients can only be determined up to certain components which are independent of the fields. These components can be projected out and covariant derivatives become again well defined. However, if one searches for a generalisation of the Riemann and Ricci tensor as well as of the Ricci scalar one cannot find a non-zero form that satisfies all the $4$ requirements of [16] i.e. 
\begin{itemize}
\item To be a tensor under the $O(D,D)$ group
\item To remain a tensor under generalised diffeomorphisms
\item to be able to obtain from the Riemann curvature, both the Ricci tensor and scalar upon suitable contractions
\item to be completely determined by the physical fields $\mathcal{H}_{MN}$ and $\phi$. 
\end{itemize}
Indeed, the components of the Riemann tensor fully determined in terms of the physical fields vanish identically as a consequence of an algebraic Bianchi identity. While it is possible to define a Riemann tensor and the resulting Ricci tensor and scalar, those are not completely determined by the physical fields. Therefore, the first effect one notices when performing such a calculation is that the counter-term required for the classical part of the entropy suffers a series of reductions. First we obtain the regularisation by taking a cutoff in the bulk manifold $\mathcal{M}$ at $\rho=\epsilon$. The resulting bulk manifold will be named $\mathcal{M}_{\epsilon}$ with the regulated conformal boundary $\partial\mathcal{M}_{\epsilon}$. The minimal surface subject to this cutoff will be called $\Sigma_{\epsilon}$. The boundary of the minimal surface in the cutoff manifold will be named $\partial\Sigma_{\epsilon}$. There will be three extrinsic curvatures emerging in this context: 
\begin{equation}
\begin{array}{ccc}
\Sigma_{\epsilon}\hookrightarrow \mathcal{M}_{\epsilon}, & \partial\Sigma_{\epsilon}\hookrightarrow \Sigma_{\epsilon}, &\partial\Sigma_{\epsilon}\hookrightarrow \partial\mathcal{M}_{\epsilon}\\
\end{array}
\end{equation}
For counter-terms only $\partial\Sigma_{\epsilon}\hookrightarrow \partial\mathcal{M}_{\epsilon}$ is relevant as it is intrinsic to the regulated boundary. As $\mathcal{R}_{ab}$ and $\mathcal{R}$ refer to the intrinsic curvature of the boundary of the minimal surface and $\mathcal{K}$ and $\mathcal{K}_{ab}$ refers to the extrinsic curvature of the minimal surface embedded into the cutoff surface, our cutoff term will be simplified. However, due to Gauss-Codazzi relations of the form $\mathcal{R}=\mathcal{K}^{2}-\mathcal{K}_{ab}\mathcal{K}^{ab}$, the extrinsic curvature enters implicitly. 
\\
\section{7. Doubled space connections and curvatures}
In the classical approach it was only the curvature intrinsic to the regulated boundary that played a role in the description of the counter-term. In the double field theoretical context we have isometries over the entire bulk space which can be continued into the boundary. The main result, as shown in [16] is that all these curvatures can either be arbitrary, when depending on parameters not related to the physical fields, or, become null when the undetermined connection components drop out from traces of curvatures. Only by applying projection operators as will be seen shortly will the entanglement entropy counter-term formulas make sense in the context of double field theory. 

As has been shown in [16], the existence of a generalised Riemann tensor $\mathcal{R}_{MNPQ}$ can be proved. The duality covariant generalised Riemann tensor determines both $\mathcal{R}_{MN}$ and $\mathcal{R}$. However, $\mathcal{R}_{MNPQ}$ is not fully determined by the physical fields. Moreover, the components of $\mathcal{R}_{MNPQ}$ that do not contain undetermined connections are zero. Indeed the generalised metric is constrained. Therefore we must introduce projectors according to [25]
\begin{equation}
\begin{array}{cc}
P_{M}^{\;\;\; N}=\frac{1}{2}(\delta_{M}^{\;\;\; N}-\mathcal{H}_{M}^{\;\;\;N}),& \bar{P}_{M}^{\;\;\; N}=\frac{1}{2}(\delta_{M}^{\;\;\; N}+\mathcal{H}_{M}^{\;\;\; N})\\
\end{array}
\end{equation}
with the properties that $P+\bar{P}=1$, $P\bar{P}=0$, $P^{2}=P$, and $\bar{P}^{2}=\bar{P}$. They represent projections on the left-handed or right-handed subspaces. 
I will follow reference [16] in using the following notation for projected indexes
\begin{equation}
\begin{array}{c}
W_{\underline{M}}=P_{M}^{\;\;\; N}W_{N}\\
\\
W_{\bar{M}}=\bar{P}_{M}^{\;\;\; N}W_{N}
\end{array}
\end{equation}

In general relativity demanding that the metric is covariantly constant fixes the connection and determines the Christoffel symbols. In double field theory however, the metric incorporates not only the spacetime metric $g_{\mu\nu}$ but also the two-form gauge field $B_{\mu\nu}$ appearing in the action through the three-form field strength $H_{\lambda\mu\nu}$, and the scalar dilaton $\phi$. The symmetries demanded from such a theory will have to incorporate diffeomorphism invariance, one-form gauge symmetry, and T-duality [15]. Being defined in the doubled space, the differential operator $\nabla_{C}=\partial_{C}+\Gamma_{C}$ acts on a generic object with $O(D,D)$ indexes as 
\begin{widetext}
\begin{equation}
\nabla_{C}T_{A_{1}...A_{n}}=\partial_{C}T_{A_{1}...A_{n}}-\omega\Gamma^{B}_{\;\;BC}T_{A_{1}...A_{n}}+\sum_{i=1}^{n}\Gamma_{CA_{i}}^{\;\;\;\;\;\;B}\;T_{A_{1}...A_{i-1}BA_{i+1}...A_{n}}
\end{equation}
\end{widetext}
where $\omega$ represents the weight defined according to [15] for each field (here only the dilaton has nontrivial weight $\omega=1$). The connection satisfies 
\begin{equation}
\begin{array}{cc}
\Gamma_{CAB}+\Gamma_{CBA}=0,& \Gamma_{ABC}+\Gamma_{CAB}+\Gamma_{BCA}=0\\
\end{array}
\end{equation}
However, we need additional conditions involving the projection operators. We define the double field theory dilaton $d$ as
\begin{equation}
e^{-2d}=\sqrt{-g}e^{-2\phi}
\end{equation}
Requiring covariant constance implies
\begin{equation}
\begin{array}{cc}
\nabla_{A}P_{BC}=0,& \nabla_{A}\bar{P}_{BC}=0\\
\end{array}
\end{equation}
\begin{equation}
\begin{array}{c}
\nabla_{A}d=\partial_{A}d+\frac{1}{2}\Gamma^{B}_{\;\;BA}=0\\
\end{array}
\end{equation}
The derivative defined in [15] is in general not double gauge covariant. Its defining property is that it generates $O(D,D)$ and double gauge covariant quantities when combined with the projections defined above, e.g. 
\begin{equation}
P_{C}^{\;\;D}\bar{P}_{A_{1}}^{\;\;B_{1}}\bar{P}_{A_{2}}^{\;\;B_{2}}...\bar{P}_{A_{n}}^{\;\;B_{n}}\nabla_{D}T_{B_{1}B_{2}...B_{n}}
\end{equation}
The derivative defined as above has been called by [15] "semi-covariant".
Imposing such covariant constraints only determines a part of the connections. As shown in [16], when we keep all the connection components, even those not determined by physical fields we obtain proper connections and fully covariant expressions. However, for the projections in which the undetermined connection components are eliminated, our curvatures become systematically zero. 
Indeed it has been shown in [16] that $\mathcal{R}_{MN}^{\;\;\;\;\;\;MN}=0$  and in order to obtain the scalar curvature leading to our double action functional, we have to contract the fully projected tensors yielding 
\begin{equation}
\mathcal{R}=\mathcal{R}^{\underline{M}\;\underline{N}}_{\;\;\;\;\;\;\;\underline{M}\;\underline{N}}
\end{equation}
The generalised tensor used in this way is however not fully determined in terms of the physical fields [16]. 
The full Christoffel connection as determined in [16] can be written as 
\begin{equation}
\Gamma_{MNK}=\hat{\Gamma}_{MNK}+\Sigma_{MNK}
\end{equation}
where $\Sigma_{MNK}$ is the part of the connection not determined by physical fields. Indeed it depends on over- and underlined indices 
\begin{equation}
\Sigma_{MNK}=\tilde{\Gamma}_{\underline{M}\;\underline{N}\;\underline{K}}+\tilde{\Gamma}_{\bar{M}\bar{N}\bar{K}}
\end{equation}
The entanglement entropy has been written in [23] in a form that makes its renormalisation counter-terms follow precisely those appearing in a quantum field theory action functional. The effective theory has been written for the double field theory in terms of the generalised curvature $\mathcal{R}$ and hence the invariant action in double field theory can be written as 
\begin{widetext}
\begin{equation}
S=\int dx\cdot d\tilde{x}\cdot e^{-2\phi}\mathcal{R}=\int dx\cdot d\tilde{x}\cdot e^{-2\phi}\mathcal{R}_{\underline{M}\;\underline{N}}^{\;\;\;\;\;\;\;\underline{M}\;\underline{N}}=\int dx\cdot d\tilde{x}\cdot e^{-2\phi}P^{MK}P^{NK}\mathcal{R}_{MNKL}
\end{equation}
\end{widetext}
which makes the undetermined pieces of the connection drop out. 
The variation of $\mathcal{R}_{MNKL}$ due to a finite variation of the connection $\Gamma\rightarrow\Gamma+\delta\Gamma$ has been calculated in [16]
\begin{widetext}
\begin{equation}
\mathcal{R}_{MNKL}(\Gamma+\delta \Gamma)=\mathcal{R}_{MNKL}(\Gamma)+2\nabla_{[M}\delta\Gamma_{N]KL}+2\Gamma_{[MN]}^{\;\;\;\;\;\;\;\;P}\delta\Gamma_{PKL}+2\delta\Gamma_{[M|QL|}\delta\Gamma_{N]K}^{\;\;\;\;\;\;\;Q}
\end{equation}
\end{widetext}
Starting with the determined connection components and considering the variation by an undetermined one, following [16] one obtains 
\begin{widetext}
\begin{equation}
\mathcal{R}_{MNKL}=\hat{\mathcal{R}}_{MNKL}+2\hat{\nabla}_{[M}\Sigma_{N]KL}+2\hat{\Sigma}_{[K}\Sigma_{L]MN}+2\Sigma_{[M|QL|}\Sigma_{N]K}^{\;\;\;\;\;\;\;Q}+2\Sigma_{[K|QN|}\Sigma_{L]M}^{\;\;\;\;\;\;\;Q}+\Sigma_{AMN}\Sigma^{Q}_{\;\;\;KL}
\end{equation}
\end{widetext}
where hatted components depend on the determined connection components.
The freedom of introducing non-physical elements is particularly important for counter-terms. By means of the replica trick in the undoubled theory, ref. [25] computed the counter-terms for the asymptotically locally $AdS_{D+2}$ spacetime resulting in relations depending on the Ricci tensor and scalar curvature
\begin{widetext}
\begin{equation}
S_{ct}=\frac{1}{16\pi G_{D+2}}\int_{\partial\mathcal{M}}d^{D+1}x\cdot \sqrt{h}\cdot (2D+\frac{1}{(D-1)}\mathcal{R}+\frac{1}{(D-3)(D-1)^{2}}(\mathcal{R}_{\mu\nu}\mathcal{R}^{\mu\nu}-\frac{D+1}{4D}\mathcal{R}^{2})+...)
\end{equation}
\end{widetext}
In double field theory the full connection depending on undetermined components will result in vanishing Riemannian curvature, hence such terms will have to be regarded in the sense of projected curvatures. In this case the undetermined connections have been shown in [16] to drop out. 

 It must be noted that the components of the Riemann tensor that are fully determined in terms of the physical fields vanish identically due to an algebraic Bianchi identity. What we may use in defining the entropy counter-terms however is a $O(D,D)$ generalised tensor which determines the Ricci tensor and the Ricci curvature but which is not fully determined in terms of the physical fields. 
 Making an order by order analysis there will remain components undetermined by the physical fields in the undoubled theory unless the constraint imposed by the projection operators is used. 

\section{8. Light sheets in double field theory}
A main difficulty in understanding the cosmological expansion in terms of the holographic principle is the fact that light-sheets in a de-Sitter spacetime do not have a clear non-positive expansion. By introducing T-duality in the bulk theory and defining a double field theory however, we obtain Ricci tensors and scalars that are not directly constrained by physical considerations. The main argument in this article is that such freedom may be employed to renormalise the entanglement entropy in an universal sense by explicitly manifesting T-duality in the bulk. When T-duality is used and a double field theory is employed, we cannot restrict ourselves to an anti-de-Sitter space, as the de-Sitter space appears as its dual partner. This apparent complication however has several beneficial features. When calculating the entanglement entropy from the bulk perspective the main idea was to identify the bulk light-sheets that correspond to the boundary light-sheets, and essentially to extend the boundary light-sheets in a consistent manner inside the bulk. As long as the bulk space was anti-de-Sitter this was immediately possible even in a covariantly meaningful sense, as shown in [30] and many subsequent articles. If we consider however the de-Sitter space in the bulk on its own, the construction of light-sheets with non-positive expansion becomes non-natural. T-duality and double field theory solve this issue in an elegant fashion. Indeed, by employing T-duality it becomes clear that the bulk cannot be considered only as anti-de-Sitter. The T-dual partner spacetime is de-Sitter and is explicitly taken into account when the extended metric is being used. However, not only does the de-Sitter space in this context preserve the proper holographic interpretation, but it also provides the required terms for a consistent renormalisation of entanglement entropy. To write an extension of the holographic entanglement entropy calculation for a bulk spacetime with manifest T-duality we must go back to the definition of light-sheets and congruences. The congruence of time-like geodesics has a tangent vector field which I will note with $\xi^{a}$. Its normalisation relation implies $\xi^{a}\xi_{a}=1$. We can then construct the $(0,2)$ tensor $\Xi_{ab}=\nabla_{b}\xi_{a}$ which satisfies 
\begin{equation}
\begin{array}{cc}
\xi^{a}\Xi_{ab}=0, & \xi^{b}\Xi_{ab}=0\\
\end{array}
\end{equation}
The first relation results from the normalisation condition of $\xi^{a}$ while the second relation is the geodesic equation. We can define a projector tensor $h_{ab}=g_{ab}+\xi_{a}\xi_{b}$ which projects all tensors in a subspace orthogonal to $\xi^{a}$. There exists a foliation of the manifold by hypersurfaces orthogonal to $\xi^{a}$. The metric that $g_{ab}$ induces on these hypersurfaces is exactly $h_{ab}$. $\Xi_{ab}$ has no components in the direction of $\xi^{a}$ and hence $\Xi_{ab}$ is defined in the subspace orthogonal to $\xi^{a}$. This subspace is the space of hypersurfaces which foliate the manifold when $\xi^{a}$ is orthogonal to the foliating hypersurfaces. Given a congruence we can compute its expansion by means of the Raychaudhuri equation [31] as
\begin{equation}
\frac{d\theta}{d\tau}=-\frac{1}{3}\theta^{2}-\sigma_{ab}\sigma^{ab}+\omega_{ab}\omega^{ab}-R_{ab}\xi^{a}\xi^{b}
\end{equation}
where 
\begin{equation}
\begin{array}{c}
\theta=\Xi^{ab}h_{ab}\\
\\
\sigma_{ab}=\Xi_{(ab)}-\frac{1}{3}\theta h_{ab}\\
\\
\omega_{ab}=\Xi_{[ab]}\\
\end{array}
\end{equation}
which are the trace, the trace-free symmetric part and the trace-free antisymmetric part of $\Xi_{ab}$ calculated with respect to the metric $h_{ab}$. They are also known as expansion, $\theta$, shear, $\sigma_{ab}$, and twist, $\omega_{ab}$ of the congruence. This means $\Xi_{ab}$ can be decomposed as 
\begin{equation}
\Xi_{ab}=\frac{1}{3}\theta h_{ab}+\sigma_{ab}+\omega_{ab}
\end{equation}
By Raychaudhuri's equation, we can see that the equation for the expansion of the congruence is determined, aside from the twist and shear, by the Ricci tensor through the term $-R_{ab}\xi^{a}\xi^{b}$. $\tau$ is an affine parameter of the geodesics which can be taken as the proper time of an observer moving along a geodesic within the congruence. 
In the case of congruences of null geodesics the construction of the transverse parts of the deviation vector and the spacetime metric become non-trivial. We may assume an affine parametrisation in the sense of $dx^{a}=k^{a}d\lambda$ with $k^{a}k_{a}=0$ and $k^{a}\xi_{a}=0$ where $\xi^{a}$ is now the deviation vector. If we write again $h_{ab}=g_{ab}+k_{a}k_{b}$ in order to obtain the induced metric on the orthogonal surfaces we notice that $k^{a}h_{ab}\neq 0$ and hence we cannot construct the transverse metric in the same way. Moreover, such a tensor would not play the role of a projector on the transverse hypersurfaces. In order to construct a meaningful transverse metric we introduce an auxiliary null vector $N^{a}$ such that $k_{a}N^{a}=-1$. We can choose $k_{a}=-\partial_{a}u$, $(u=t-x)$ and we have $N_{a}=-\frac{1}{2}\partial_{a}v$ and hence $h_{ab}=g_{ab}+k_{a}N_{b}+k_{b}N_{a}$. This will now satisfy $k^{a}h_{ab}=0$ and $N^{a}h_{ab}=0$ as it should. The Raychaudhuri equation becomes then
\begin{equation}
\frac{d\theta}{d\lambda}=-\frac{1}{2}\theta^{2}-\sigma_{ab}\sigma^{ab}+\omega_{ab}\omega^{ab}-R_{ab}k^{a}k^{b}
\end{equation}
where now the tangent vector fields become null tangent vector fields and the congruence will be one of null geodesics. The affine parameter is defined as the one which preserves the geodesic equation in its original form. The subspace considered is now two-dimensional and hence the factor arising in front of the expansion is $1/2$. 
All terms of the equation will be relevant in string theory. At this point however, I will only focus on the Ricci tensor part as it is a relevant component in the entanglement entropy counter-terms. 
Doubling the degrees of freedom changes the vector fields in the usual way presented previously. We consider extending the light-sheet formed by this type of congruences within the bulk. The Ricci tensor now will be written in terms of doubled coordinates and it will be identically zero whenever only physical degrees of freedom are considered in order to determine it. In order to define the Ricci tensor in a meaningful way, we need non-physical degrees of freedom. However, this fact is not surprising if one considers that whenever T-duality is manifest, as is the case in double field theory, both AdS and dS spaces are taken into consideration as dS is the T-dual of AdS. The definition $\Xi_{ab}=\nabla_{b}\xi_{a}$ must not only take into account the generalisation to doubled coordinates and hence become $\Xi_{AB}=\nabla_{B}\xi_{A}$, but must also take into account the generalised connection and covariant derivative induced by the manifest presence of T-duality in the bulk. Using the modified covariant derivative in the bulk 
\begin{equation}
\nabla_{A}\rightarrow \rhd_{A}=\nabla_{A}+\frac{1}{2}J^{B}(\eta_{BA}+B_{BA})
\end{equation}
and using in the derivation of the light-sheet this expression, we notice several modifications. First, our orthogonal tensor $\Xi_{AB}=\nabla_{B}k_{A}$ will acquire additional stringy terms in the form of 
\begin{equation}
\Xi_{AB}=\rhd_{B}k_{A}=\nabla_{B}k_{A}+\frac{1}{2}J^{C}\eta_{BC}k_{A}+\frac{1}{2}J^{C}B_{BC}k_{A}
\end{equation}
where $B_{BC}$ represents the $B$-field and $J^{C}$ represents the left invariant current. The doubled metric is the one combining de-Sitter and anti-de-Sitter indices. The Raychaudhuri formula for the expansion will also receive corrections due to the doubled bulk geometry, including de-Sitter components which will alter the overall Ricci tensor
\begin{widetext}
\begin{equation}
\frac{d\theta}{d\lambda}=-\frac{1}{d-2}\theta^{2}-\sigma_{AB}\sigma^{AB}+\omega_{AB}\omega^{AB}-(R_{AB}^{(AdS)}+R_{AB}^{(dS)})k^{A}k^{B}
 \end{equation}
 \end{widetext}
 where aside of the modifications in the shear and rotation terms due to stringy corrections, I manifestly introduced de-Sitter components in the Ricci tensor.
As the Raychaudhuri equation is a geometrical identity, it is clear that stringy corrections to the light-sheet geometry arise, having as an effect the alteration of the light-sheet expansion. Moreover, the orthogonal foliation of the doubled light-sheet will also receive stringy (doubled field) corrections.
The calculation of the entanglement entropy however depends on the minimal area and the counter-terms needed for a consistent renormalisation of the entanglement entropy depend on the Ricci tensor. 
It has been shown in [16] that while a T-duality covariant generalised Riemann tensor exists that determines both the Ricci tensor and the Ricci scalar, such a tensor is not fully determined in terms of physical fields. This arises as a consequence of an algebraic Bianchi identity [16]. The components of the Riemann tensor that do not contain undetermined connections however are identically zero. Meaningful Ricci tensors can be derived in double field theory by taking into account suitable projections. Similar constraints can be introduced starting with the mixing of de-Sitter and anti-de-Sitter Ricci tensors. 
\section{9. Extremal area in doubled AdS bulk space}
At this point we have all the required tools to start computing the analogue of the minimal area in the Ryu-Takayanagi formula. It is important to underline the role of the additional stringy terms in the renormalisation procedure as well as the appearance of extra terms that would correct the standard Ryu-Takayanagi formula. It has been shown in reference [32], [33], and [34] that T-duality connects de-Sitter and anti-de-Sitter spacetimes. Particularly applying a time-like T-duality to type $II$ string theory results in the so called [35] type $II^{*}$ string theory which leads to an effective theory with a modified sign for the cosmological constant. The anti-de-Sitter and the de-Sitter metrics are given by 
\begin{equation}
\begin{array}{cc}
Ads: & \frac{d\rho^{2}}{4\rho^{2}}-\frac{1}{\rho}dt^{2}+\frac{1}{\rho}\delta_{ij}dx^{i}dx^{j}\\
\\
dS: & -\frac{d\rho^{2}}{4\rho^{2}}-\frac{1}{\rho}dt^{2}+\frac{1}{\rho}\delta_{ij}dx^{i}dx^{j}
\end{array}
\end{equation}
where in the case of the anti-de-Sitter space we consider the near horizon limit while in the de-Sitter space  we consider the near light-cone limit. Trying to discuss the holographic duality solely in terms of a de-Sitter spacetime fails systematically, mainly because there is no obvious suitable dual conformal field theory. If such a theory existed, it is assumed to be Euclidian and non-unitary. Moreover, the de-Sitter space alone is difficult to realise as a vacuum space of string theory. However, in the context of double field theory, de-Sitter and anti-de-Sitter spacetimes appear together, representing each the solution of the problems arising in the other. Indeed, the appearance of the T-dual de-Sitter spacetime naturally provides a connection between the counter-terms of the renormalisation procedure of the entanglement entropy as calculated in a holographic context. Of course, once double field theory is considered in the bulk, the boundary theory will generalise to one with an extended metric that will have negative signs, hence in general we will not have an Euclidian signature on the boundary except for the limit cases when we ignore the doubled structure in the bulk. Also, non-unitarity results from the fact that the eigenvalues of the conformal dilatation operators $L_{0}$ and $\bar{L}_{0}$  become complex and hence the energy eigenvalues become complex. This problem may also be solved by manifestly introducing T-duality and treating the de-Sitter and the anti-de-Sitter spacetimes as dual pairs in an extended theory. I will leave the discussion of these aspects for a future article. Here, I just note that extending the theory to include T-duality has an impact on the computation of the counter-terms for entanglement entropy and provides tools for a better understanding of the holographic aspects of both de-Sitter and anti-de-Sitter spacetimes. It has been noted in ref. [36] that the mapping between a Lorenzian bulk space and an Euclidean boundary would imply changes both in the central charge of the CFT algebra and in the conformal generators. Despite of these, what appears to remain unchanged is the entropy as calculated by means of a naive application of the Cardy formula. While certainly, the calculation of [36] implies a non-unitary theory due to the complex energy eigenvalues, it appears that the terms combine in such a way that they provide a meaningful entropy. As both de-Sitter and anti-de-Sitter entropies need counter-terms of similar forms, bringing them together by manifestly introducing T-duality leads to a series of simplifications. The fact that the entropy calculation in the context of conical defects on de-Sitter space [36] remains meaningful, combined with the observation that T-duality relates de-Sitter and anti-de-Sitter spaces suggests that full renormalisation of entanglement entropy is a well defined concept in string theory and that a string theoretical description must combine de-Sitter and anti-de-Sitter spacetimes. 
Let us first derive the counter-terms for the entanglement entropy in the AdS spacetime. First, following [23], consider the entangling surface $\Sigma$ as a codimension $2$ surface of the bulk $\mathcal{M}$. The coordinates on $\Sigma$ are $(\rho, x^{a})$ with $a=1,...,D-1$. In order to embed $\Sigma$ into $\mathcal{M}$ we introduce the coordinates 
\begin{equation}
X^{m}=(\rho,t,x^{1},...,x^{D-1},y(\rho,x^{a}))
\end{equation}
taking the time coordinate $t$ constant. The regularisation of the bulk means the restriction to $\mathcal{M}_{\epsilon}$ with $\rho\geq \epsilon > 0$. The regulated entangling surface will then be $\Sigma_{\epsilon}$. The surface $\Sigma_{\epsilon}$ is a constant time hypersurface of $\mathcal{M}_{\epsilon}$. The metric $\gamma_{\alpha\beta}$ on $\Sigma_{\epsilon}$ then is 
\begin{widetext}
\begin{equation}
ds_{\Sigma_{\epsilon}}^{2}=(\frac{1}{4\rho^{2}}+\frac{1}{\rho}(\partial_{\rho}y)^{2})d\rho^{2}+\frac{2}{\rho}\partial_{\rho}y\cdot \partial_{a}y\cdot dx^{a}+\frac{1}{\rho}(\delta_{ab}+\partial_{a}y\cdot \partial_{b}y)\cdot dx^{a}\cdot dx^{b}
\end{equation}
\end{widetext}
The regulated bare entanglement entropy in this case becomes
\begin{widetext}
\begin{equation}
S_{reg}=\frac{1}{4G_{D+2}}\int_{\partial\Sigma}d^{D-1}\int_{\epsilon}^{\rho}d\rho\frac{1}{2\rho^{(D+1)/2}}\sqrt{1+4\rho(\partial_{\rho}y)^{2}+(\partial_{a}y)^{2}}
\end{equation}
\end{widetext}
We may now expand $y(\rho,x^{a})=y^{(0)}+y^{(1)}\rho+...$ with 
\begin{equation}
y^{(1)}=\frac{1}{2(D-1)}(\partial_{a}\partial^{a}y^{(0)}-\frac{\partial_{a}y^{(0)}\cdot \partial_{a}\partial_{b}y^{(0)}\cdot\partial_{b}y^{(0)}}{1+(\partial_{c}y^{(0)})^{2}})
\end{equation}
With the asymptotic expansion near the conformal boundary we obtain the regulated expression for $D=3$ as 
\begin{widetext}
\begin{equation}
S_{reg}=\frac{1}{4G_{5}}\int_{\partial\Sigma_{\epsilon}}dx^{2}(1+(\partial_{c}y^{(0)})^{2})^{1/2}(\frac{1}{2\epsilon}-\frac{\partial_{a}y^{(0)}\cdot \partial_{a}y^{(1)}+2y^{(1)2}}{2(1+(\partial_{b}y^{(0)})^{2})}\cdot log(\epsilon)+...)
\end{equation}
\end{widetext}
and for $D>3$ as
\begin{widetext}
\begin{equation}
S_{reg}=\frac{1}{4G_{D+2}}\int_{\partial\Sigma_{\epsilon}}d^{D-1}x(1+(\partial_{c}y^{(0)})^{2})^{1/2}(\frac{\epsilon^{-\frac{(D-1)}{2}}}{D-1}+\frac{\epsilon^{-\frac{(D-3)}{2}}}{D-3}\cdot \frac{\partial_{a}y^{(0)}\cdot \partial_{a}y^{(1)}+2y^{(1)2}}{1+(\partial_{b}y^{(0)})^{2}}+...)
\end{equation}
\end{widetext}
Finding now counter-terms that are integrals over covariant quantities defined on $\partial\Sigma_{\epsilon}$ has been explained in [23]. The induced metric on $\partial\Sigma_{\epsilon}$, $\tilde{\gamma}_{ab}$ is given by 
\begin{equation}
ds_{\tilde{\gamma}}^{2}=\frac{1}{\epsilon}(\delta_{ab}+\partial_{a}y\cdot \partial_{b}y)dx^{a}dx^{b}
\end{equation}
with determinant 
\begin{equation}
\tilde{\gamma}=det(\tilde{\gamma}_{ab})=\epsilon^{-\frac{D-1}{2}}(1+\partial_{a}y^{2})
\end{equation}
expanding the volume form as
\begin{widetext}
\begin{equation}
\sqrt{\tilde{\gamma}}=\epsilon^{-\frac{D-1}{2}}(1+(\partial_{c}y^{(0)})^{2})^{1/2}(1+\epsilon\frac{\partial_{b}y^{(0)}\cdot \partial_{b}y^{(0)}}{1+(\partial_{c}y^{(0)})^{2}}+...)
\end{equation}
\end{widetext}
Therefore, the leading divergence in the entanglement entropy comes from an area divergence and hence the first counter-term in the AdS case is 
\begin{equation}
S_{ct,1}=-\frac{1}{4G_{D+2}}\frac{1}{D-1}\int_{\partial\Sigma_{\epsilon}}d^{D-1}x\sqrt{\tilde{\gamma}}
\end{equation}
The counter-terms for the sub-leading divergences can be obtained for $D=3$ considering 
\begin{widetext}
\begin{equation}
S_{reg}+S_{ct,1}=-\frac{1}{8G_{5}}\int_{\partial\Sigma_{\epsilon}}d^{2}x\sqrt{\tilde{\gamma}}\frac{y^{(1)2}}{1+(\partial_{c}y^{(0)})^{2}}\cdot log(\epsilon)+...
\end{equation}
\end{widetext}
This expression must be written in a covariant way, by noticing that on a constant time hypersurface of the regulated boundary the metric is 
\begin{equation}
ds^{2}_{D}=\tilde{g}_{ij}dx^{i}dx^{j}=\frac{1}{\epsilon}\delta_{ij}dx^{i}dx^{j}
\end{equation}
Taking the embedding of $\partial\Sigma_{\epsilon}$ given by $X^{D}=y(\epsilon, x^{a})$, the unit normal co-vector is 
\begin{equation}
n=\epsilon^{-\frac{1}{2}}(1+\partial_{c}^{2})^{-\frac{1}{2}}(\partial_{a}y\cdot dx^{a}-dx^{D})
\end{equation}
and we define the induced metric $\tilde{\gamma}_{ij}$ and the extrinsic curvature $K_{ij}$ as
\begin{equation}
\begin{array}{cc}
\tilde{\gamma}_{ij}=\tilde{g}_{ij}-n_{i}n_{j},& K_{ij}\tilde{\gamma}_{i}^{k}\nabla_{k}n_{j}
\end{array}
\end{equation}
where the covariant derivative with respect to $\tilde{g}_{ij}$ is $\nabla_{k}$. The scalar extrinsic curvature will then be
\begin{widetext}
\begin{equation}
 K=\frac{\epsilon^{\frac{1}{2}}}{(1+(\partial_{c}y)^{2})^{\frac{1}{2}}}(\partial_{a}\partial^{a}y-\frac{\partial_{a}y\cdot \partial_{a}\partial_{b}y\cdot \partial_{b}y}{1+(\partial_{e}y)^{2}})=2(D-1)\frac{\epsilon^{\frac{1}{2}}y^{(1)}}{(1+(\partial_{c}y^{(0)})^{2})^{\frac{1}{2}}}+...
 \end{equation}
\end{widetext}
The Ricci scalar will have the form 
\begin{widetext}
\begin{equation}
R=\frac{2\epsilon}{(1+(\partial_{e}y)^{2})}((\partial_{a}\partial^{a}y)^{2}-\partial_{a}\partial_{b}y\cdot \partial_{a}\partial_{b}y-\partial_{a}y\cdot \partial_{b}y\cdot \frac{\partial_{a}\partial_{b}y\cdot \partial_{c}\partial^{c}y-\partial_{a}\partial_{c}y\cdot \partial_{b}\partial_{c}y}{1+(\partial_{d}y)^{2}})
\end{equation}
\end{widetext}
Using these, we find for $D=3$ the logarithmic counter-term to be 
\begin{equation}
S_{ct,2}=\frac{1}{64 G_{5}}\int_{\partial \Sigma_{\epsilon}}d^{2}x \sqrt{\tilde{\gamma}}K^{2}\log(\frac{\epsilon}{\mu})
\end{equation}
where $\mu$ is the cutoff scale. For higher dimensions $D>3$ we have
\begin{equation}
S_{ct,2}=-\frac{1}{8G_{D+2}}\frac{1}{(D-1)^{2}(D-3)}\int_{\partial\Sigma_{\epsilon}}d^{D-1}x\sqrt{\tilde{\gamma}}K^{2}
\end{equation}
In the dual case, when de-Sitter space is considered the situation is slightly different. However, we find the same type of divergencies and hence the counter-terms defined above can be identified with the terms arising in the context of double field theory as divergencies with opposite sign. The two formulations are related by means of T-duality and hence a complete T-duality invariant theory must contain both of them. It therefore appears that the counter-terms required for the consistent renormalisation of the entanglement entropy in the anti-de-Sitter context are given precisely by the T-dual counter-part. Moreover, as T-duality may in certain situations alter the topology of the underlying spacetime, it is expected that topologies related by T-duality may in fact be two sides of the same coin, one cancelling the divergencies of the others. The role of string theory as a divergence remover then may be broader than suspected up to now. 
Assuming one can find translation invariance with respect to a boundary Euclidean time direction, we may construct a subdomain of a Euclidean time slice in the future boundary $\mathcal{I}^{+}$. When we separate the complement of this region we observe a loss of information that can be associated to an entropy resulting from the entanglement of the region with the complement. We may consider studying extremal surfaces in de-Sitter space on a constant boundary Euclidean time slice anchored on the subspace of the future spacelike boundary and entering the bulk in the time direction towards the past. It has been shown in [40] that in the case of de-Sitter spacetime the signs associated to the resulting terms differ from the anti de-Sitter case. Moreover if we consider only solutions given in the form of real surfaces there is no natural turning point where the surface stops dipping inward, as is the case for anti de-Sitter. Given sufficient symmetry, one can define extremal surfaces as unions of two half-extremal surfaces joined continuously but not smoothly. The requirement that the area is minimal makes the surface null and the area equal to zero. These solutions are actually just boundaries of the past light-cone wedge of the given subregion. These surfaces appear to have null entropy and do not seem to determine the entanglement entropy in any de-Sitter/CFT duality. Other extrema however may generate entanglement entropy in the form we need. Indeed, it has been shown in [40] that if we look for other extrema, for example complex saddle points, we obtain meaningful terms that can be interpreted as contributing to the entanglement entropy. There is of course a problem related to the fact that the boundary theory is Euclidean and hence the divergences will not match the ones obtained in the anti de-Sitter case, as they will be rotated towards the imaginary axis. However, when performing the regularisation we do not work on the Euclidean boundary but instead we again define a regulated de-Sitter bulk which is Lorentzian. The regularisation procedure at all steps occurs in the Lorentzian bulk and hence the cancellation of the divergences has the same structure as in the anti de-Sitter case. In three dimensions the calculation for the de-Sitter spaces diverges. We can see taking in account [41] that, if we cut time off at a large value, we obtain
\begin{equation}
S=\frac{1}{8\pi G}\int d^{2}x\cdot e^{2t/l}\frac{-1}{l}+fin
\end{equation}
where $fin$ refers to finite terms. This expression diverges for large times. The divergences can be cancelled by adding local boundary counter-terms. In three dimensions one may add the required counter-terms and one obtains
\begin{equation}
S=S_{B}+\frac{1}{8\pi G}\int_{\partial\mathcal{M}^{+}}d^{2}x\cdot \sqrt{\tilde{\gamma}}\cdot \frac{1}{l}+\frac{1}{8\pi G}\int_{\partial\mathcal{M}^{-}}d^{2}x\cdot \sqrt{\tilde{\gamma}}\cdot \frac{1}{l}
\end{equation}
with $S_{B}$ being 
\begin{widetext}
\begin{equation}
S_{B}=-\frac{1}{16\pi G}\int_{\mathcal{M}}d^{D+1}x\cdot \sqrt{-\gamma}\cdot (R+2\Lambda)+\frac{1}{8\pi G}\int^{\partial\mathcal{M}^{+}}_{\partial\mathcal{M}^{-}}d^{D}x\cdot \sqrt{\tilde{\gamma}}\cdot K
\end{equation}
\end{widetext}
where $\mathcal{M}$ is the bulk manifold and $\partial\mathcal{M}^{\pm}$ are the spatial boundaries at early and late times, $\gamma_{\mu\nu}$ is the induced metric and $K$ is the extrinsic curvature of the boundaries. In de-Sitter space the spacetime boundaries $\mathcal{I}^{\pm}$ are Euclidean surfaces at early and late time infinity. The length scale used is defined as 
\begin{equation}
l=\sqrt{\frac{D(D-1)}{2\Lambda}}
\end{equation}
It is clear that the supplemental terms have cancelled the divergences. For any dimensions one can write according to [36] the general counter-terms
\begin{equation}
S_{ct}=\frac{1}{8\pi G}\int_{\partial\mathcal{M}^{+}}d^{2}x\cdot \sqrt{\tilde{\gamma}}L_{ct}+\frac{1}{8\pi G}\int_{\partial\mathcal{M}^{-}}d^{2}x\cdot \sqrt{\tilde{\gamma}}L_{ct}
\end{equation}
where 
\begin{equation}
L_{ct}=\frac{(D-1)}{l}-\frac{l^{2}}{2(D-2)}R
\end{equation}
$R$ being the intrinsic curvature of the boundary surface. 
It is worth noting that double field theory connects any counter-term of dimension $D$ in anti de-Sitter space to a counter-term of dimension $D-1$ in de-Sitter space. 
The two counter-terms cannot be independent as T-duality relates de-Sitter and anti de-Sitter spacetimes and indeed if we look at the two formulae we see they are related, dealing with similar geometric structures (although of course their interpretation is different, as in de-Sitter spacetime we have a cosmological horizon as a future horizon and not an anti de-Sitter type enclosing surface).
Therefore, due to the manifest presence of T-duality in double field theory, the entanglement entropy counter-terms will have to obey certain restrictions which do not appear in the un-doubled theory but which are capable of systematising the renormalisation procedure. 
T-duality in entanglement entropy plays in a sense the role of the observation that the splitting of terms in "bare" and "counter-terms" is unnatural when one realises the existence of the renormalisation group. In this sense T-duality shows that all scales must collaborate in order to provide a meaningful entanglement entropy and therefore the combination of de-Sitter and anti de-Sitter contributions is more natural from the perspective of quantum gravity. 
This confirms the fact that entanglement entropy is not expected to change during a T-duality transformation and is protected by the requirement of using only properly projected components. 

As T-duality is fundamentally a topology altering duality, the prescription of topological invariance introduced in [37] becomes manifest in the construction of a renormalised entanglement entropy. Indeed, when T-duality is included, different topologies related by a T-duality transformation are physically equivalent and therefore, when they arise in the bulk spacetime can be cancelled. In fact, such terms arise naturally when double field theory is employed in the bulk, their role being to define a consistent prescription for the renormalisation of entanglement entropy.

\section{10. Conclusion}
T-duality is a characteristic feature of string theory as it cannot have a well defined interpretation in the context of point particles. However, its effects do not vanish when we go to a effective field theory. Instead they only become hidden and their effects are not commonly taken fully into account in effective theories like supergravity. This represents a problem when we wish to go to cosmological or phenomenological applications of string theory, as many physical solutions are either non-manifest or become ill defined. A solution to this problem is given by double field theory. This formulation allows the manifest presence of additional coordinates which encode precisely the string-specific phenomena attached to winding numbers. Even in the case when strong constraints are imposed, and we recover basically supergravity, we may benefit from the doubled space as we can obtain solutions that would not be visible otherwise. The best known example is the existence of a plethora of stable de-Sitter geometries that were forbidden by a series of no-go theorems in the context of simple supergravity. However, another benefit we can obtain from double field theory is a set of constraints on the entanglement entropy counter-terms so that they cancel the divergencies in a systematic way. Moreover, if we demand T-duality to be manifest and analyse the calculation of entanglement entropy from the perspective of a doubled space we obtain a relation between counter terms at all orders. This relation originates from the fact that in double field theory the only form of curvature tensor that is completely determined by physical fields has all its components equal to zero. As a result the same is valid for the Ricci tensor and the curvature scalar. Due to the fact that T-duality mixes all length scales this new relation remains manifest in all orders of the quantum correction to the entanglement entropy. In fact, T-duality imposes a relation on the counter-terms in a way similar to the relations imposed by demanding that the Slavnov-Taylor equations remain valid in the renormalisation prescription of common quantum field theories. The relation resulting from T-duality however is much more general and allows for an unexpected flexibility. In fact meaningfully defining a connection in the context of T-duality manifesting doubled spacetime requires keeping certain connection components not fully determined by the physical fields. These connection components allow for extra flexibility in defining counter-terms to entanglement entropy divergencies and absorbing them into modifications of these connections which do not have any physical impact. Otherwise, in order to have meaningful Ricci curvatures and tensors for the proper definition of the generalised counter-terms arising when we extend the counter-terms from an un-doubled quantum field theory calculation of entanglement entropy, we need to impose a set of restrictions given by covariantly constant projection operators. Such restrictions lead to relations between counter-terms at all scales. 
The observation presented in this article will prove crucial for further computations relating effects originating in string theory to phenomenology at lower energy.

\end{document}